\documentclass[12pt]{article}

\usepackage{amsfonts}
\usepackage{amsmath}
\usepackage{amscd,amssymb}
\usepackage{hyperref}
\usepackage{graphicx}
\usepackage{amsthm}

\def\g{\gamma}
\newcommand{\be}{\begin{equation}}
\newcommand{\ee}{\end{equation}}
\newcommand{\bea}{\begin{eqnarray}}
\newcommand{\eea}{\end{eqnarray}}
\let\bm=\bibitem

\def\m{\mu}
\def\n{\nu}

\def\S{\Sigma}
\def\t{\tau}
\def\th{\theta}

\def\O{\Omega}

\def\a{\alpha}

\def\del{\partial}
\def\4{{\sst{(4)}}}

\def\f{\phi}
\def\del{\partial}
\def\b{\beta}
\def\a{\alpha}
\def\f{\phi}
\def\e{\epsilon}
\let\la=\label

\let\bm=\bibitem

\begin{document}
\pagenumbering{roman}

\begin{titlepage}

\vspace{3.0cm}

\

\

\

\centerline{\Large \bf Time-Dependent AdS Backgrounds from S-Branes}
\vspace{1.5cm}

\centerline{Nihat Sadik Deger\footnote{sadik.deger@boun.edu.tr}}

\

\noindent
\centerline{Department of Mathematics, Bogazici University,}\\
\centerline{Bebek, 34342, Istanbul-Turkey}

\

\centerline{Feza Gursey Center for Physics and Mathematics,}
\centerline{Bogazici University, Kandilli, 34684, Istanbul-Turkey}

\

\vspace{1.5cm}

\centerline{\bf ABSTRACT}
\vspace{0.5cm}

We construct time and radial dependent solutions that describe p-branes in chargeless S-brane backgrounds. In particular, 
there are some new M5- and D3-branes among our solutions which have AdS limits and contain a cosmological singularity as well. 
We also find a time-dependent version of the dyonic membrane configuration in 11-dimensions by applying a Lunin-Maldacena deformation to our 
new M5-brane solution.

\vspace{2cm}

\end{titlepage}

\pagenumbering{arabic}



\section{Introduction}
One of the most profound ideas in String/M-theory is the AdS/CFT correspondence \cite{maldacena}. The original proposal of Maldacena was tested in various examples and
extended to include different cases. However, understanding of this duality in time-dependent backgrounds is still incomplete. One important 
motivation to study this 
problem is to understand cosmological singularities like big bang using holography, see for instance \cite{nojiri}, \cite{hertog}, \cite{chu}.
Since AdS spacetimes appear as near horizon geometries of single and intersecting non-dilatonic p-branes, namely M2- and M5-branes in d=11 and D3-brane in d=10
(for a review see \cite{rev}), trying to add time dependence to them is a reasonable way to address this issue. 

It is a well-known fact that one can replace the Minkowski geometry of the 
worldvolume of a p-brane with any Ricci-flat spacetime \cite{ricci1, ricci2} such as the time-dependent Kasner spacetime which is a natural choice to analyze
the cosmological singularity problem. This was first considered in \cite{kasner1}
for a D3-brane  and this set-up was studied further in several other papers such as \cite{kasner2}-\cite{kasner7}.
In \cite{frw1} a more general solution where the D3-brane worldvolume is replaced by Friedmann-Robertson-Walker (FRW) metric with a conformal factor 
was obtained and its properties were analyzed \cite{frw2}. For the M5-brane only the Kasner deformation has been investigated \cite{kasnerM5, kasnerM52}. 

In this letter we will show that the solutions mentioned in the previous paragraph are some specific examples of p-branes in chargeless S-brane 
backgrounds \cite{townsend, ohta, deform}. 
S-branes \cite{pope1, pope2, s1,s2, s3} are time-dependent analogs of p-branes and they may carry electric or magnetic charge similarly. 
However, unlike p-branes, the S-brane solution is still nontrivial when its charge vanishes. Various solutions describing S-brane intersections have been obtained 
\cite{deger, ivas, ohta1, nonstandard, chern}. Moreover, intersections between S- and p-branes were also studied \cite{mas, sp} but with the 
assumption that the S-brane has a nonzero charge. It was found in \cite{mas, sp} that solutions involving non-dilatonic p-branes were possible only with smearing in the 
transverse space, which destroys the AdS near-horizon geometry. In the next section we will demonstrate that relaxing the condition on S-brane charge 
results in a new set of solutions. In particular, we find that D3- and M5-brane solutions without smearing are allowed.
It turns out that, there are two possible FRW type time-dependent D3-brane solutions and only one of which was known before \cite{frw1}. 
The new one also has $AdS_5 \times S^5$ limit when the D3-brane worldvolume contains a hyperbolic part and a cosmological singularity as well. 
Yet, the two D3 configurations differ in the exponent of the string coupling 
at the singularity which could be important for its resolution \cite{frw2}. For the M5-brane in addition to the Kasner case \cite{kasnerM5},
we show that it is possible to have spherical or hyperbolic parts along the worldvolume directions too. The latter has an AdS limit like the D3-brane.
In section 3, we provide another example of a cosmological AdS background by appling a Lunin-Maldacena deformation \cite{lm} to our M5-brane solution.
This generates the time-dependent version of the dyonic M2-brane, M2 $\subset$ M5, of \cite{dyonic} which has $AdS_7 \times S^4$ near-horizon too. 
This deformation is actually a particular U-duality transformation and a simple formula for the deformed background was given in \cite{deformations} which 
we use here. We conclude in section 4 with some comments and possible future directions.

\section{The Solution}

The action in $d$-dimensions in the Einstein frame describing the bosonic sector of various supergravity theories containing the graviton 
$g_{MN}$, the dilaton $\phi$ coupled to the q-form field strength $F_{[q]}=dA_{[q-1]}$ with the coupling constant $a$ is given as:
\be
S=\int d^d x \sqrt{-g} \left( R- \frac{1}{2} \del_{\m}\f \del^{\m} \f - 
\frac{1}{2(q)!}e^{a \f} F^2_{[q]} \right) \, . 
\label{action}
\ee
The Chern-Simons terms are omitted since they are irrelevant for our solutions.
The field equations are:
\bea
&&R_{\m\n}=\frac{1}{2}\del_\m \f \del_\n \f +
\frac{1}{2(q)!}e^{a \f} \left( q F_{\m{\a_2}...{\a_{q}}} F_\n 
^{\, \a_2...\a_{q}} - \frac{(q -1)}{d-2}F^2_{[q]}g_{\m\n} \right) \,\, , \nonumber\\
\nonumber
&&\del_\m \left( \sqrt{-g}e^{a \f} F^{\m \n_2 ...\n_{q}} \right)=0 \,\,\,\, , \\
\label{fieldeqns}
&&\frac{1}{\sqrt{-g}} \del_\m \left( \sqrt{-g}\del^\m \f 
\right)=
\frac{a}{2(q)!}e^{a \f}F^2_{[q]} \,\,\,\, .
\eea
The field strength satisfies the Bianchi identity $\del _{[\n}F_{\m_1...\m_{q}]}=0$.

We are interested in finding solutions describing a p-brane in the presence of a chargeless S-brane without any smearing in p-brane's transverse space.
The main result of this paper is that the above equations of motion admit such a solution:
\bea
ds^2 &=& H^{-\frac{d-p-3}{d-2}} \left[ e^{2 \b t} G_{n,\sigma}^{-\frac{1}{n-1}}(-G_{n,\sigma}^{-1} dt^2 +   d \Sigma_{n,\sigma}^2) + 
\sum_{i=0}^{k-1} e^{2 b_i t} dy_i^2\right] \nonumber \\
\nonumber
&+& H^{\frac{p+1}{d-2}} e^{-\frac{2\gamma\epsilon a}{d-p-3}t}(dr^2+ r^2d\O_m^2)  \,\,\,\, ,\\
\label{solution}
\f &=& \e a \ln H + \g t  \,\,\,\, , \\
\nonumber
F_{t1...pr} &=&Q \, \textrm{*[Vol} (\O_m) ]  \,\,\textrm{(Electric)} \, , \, \hspace{0.2cm}
F_{1...m} =  Q \, \textrm{Vol} (\O_m) \,\,\textrm{(Magnetic)} \, , 
\eea
where Vol$(\O_m)$ is the volume form of the $m$-dimensional unit sphere $\O_m$ and the symbol `*' denotes the Hodge dual operation with respect to full metric.
Here $p=n+k$ and the function $H$ is the harmonic function of the (m+1)-dimensional flat transverse space
\be
\label{harmonic}
H =  1 + \frac{Q}{r^{m-1}} \,\,\,\, .   
\ee
For the  D3-brane its 5-from field strength should be self-dual. In the metric above,  $d\Sigma_{n,\sigma}^2$ represents the metric on the $n$-dimensional 
unit hyperbola, unit sphere or flat space and
the function $G_{n,\sigma}$ is defined respectively as:
\be
G_{n,\sigma} =
\left\{
\begin{array}{ccc}
M^{-2}\sinh^2\left[(n-1)M\,(t-t_0)\right],
 \,\,\, &\sigma=-1& \,\,\, \textrm{(hyperbola)}\, , \\
M^{-2}\cosh^2\left[(n-1)M\,(t-t_0)\right],
 \,\,\, &\sigma=1& \,\,\, \textrm{(sphere)}\, , \\
\exp[{2(n-1)M\,(t-t_0)}], \,\,\,  &\sigma=0& 
\,\,\, \textrm{(flat)} \, .
\end{array}
\right. 
\label{G}
\ee
Finally, constants should satisfy
\bea
\nonumber
\b   &=& -\frac{1}{n-1}\left(\sum_{i=0}^{k-1} b_i - \frac{(m+1)\gamma\e a}{d-p-3}\right)   \, , \\
 n(n-1) M^2 &=& (n-1) \b^2 + \sum_{i=0}^{k-1} b_i^2 + +   \frac{(m+1) \gamma^2 a^2}{(d-p-3)^2} + \frac{\g^2}{2} \, . 
\label{restriction}
\eea
It is understood that when $k=0$ then, all \, $b_i$'s are zero. We should also have\footnote{When $\sigma=0$, having different time-dependent exponentials 
for $\S_{n,0}$ directions is also possible.} $ n\geq 2$ and $m \geq 2$.
The constant $Q$ is the charge of the p-brane and $t_0$ is another constant related to the S-brane\footnote{It is possible to add more additive constants in the 
exponentials multiplying $dy_i^2$ in the metric.}. 

If we set all time-dependent parts to zero\footnote{Note that this enforces $\sigma$=0, since when $\sigma \neq 0$ there are inverse powers of the constant $M$ in 
(\ref{G}).}, then this solution is nothing but the
well-known p-brane solution, for a review see \cite{rev}. The dilaton coupling '$a$' is zero in 11-dimensional supergravity and in type $IIA$ and $IIB$ 
supergravities it is given as:
\be
\begin{cases} \e a=\frac{3-p}{2} \, ,  \hspace{1cm}
\textrm{$RR$-branes} \cr
\e a=\frac{p-3}{2} \, ,  \hspace{1cm}
\textrm{$NS$-branes}
\end{cases}
\ee
where $\e=1$ for electric branes ($p=0,2$  in type $IIA$ 
and 
$p=1$ in type 
$IIB$) and $\e=-1$ for magnetic branes ($p=4,6$ in type 
$IIA$ 
and $p=5$ in type $IIB$). 

On the other hand, if we remove all radial dependence by setting the p-brane charge to zero, i.e. $Q=0$,
then what is left is the vacuum S-brane solution \cite{townsend, ohta, deform}. Therefore, our new solution (\ref{solution})  can be thought of as a 
superposition of a p-brane with a chargeless S-brane. 

Although, intersections between S- and p-branes were studied before in \cite{mas, sp}, this solution was missed because in those papers S-brane 
was assumed to have a charge to begin with. The intersection dimension and the amount of smearing for the p-brane  depends crucially on the 
existence of the S-brane charge and taking the zero charge limit at the end does not reproduce the above solution. It turns out that, in 
almost all the solutions found in \cite{mas, sp} there has to be some smearing \footnote{Even when there is no smearing, the zero charge limit 
only gives the $\sigma=0$ case with all $b_i$'s equal.}. In particular, this is the case for all non-dilatonic p-branes 
which we now demonstrate explicitly.
In such configurations 
the dilaton and the field strenghts are written as summations of those corresponding to S- and p-branes with appropriate $r$ and $t$ dependent 
coefficients and in the metric multiplicative separation of variables is assumed:
\bea
\label{metric}
ds^2=- e^{2A(t)} e^{2\a (r)} dt^2 &+& e^{2B(t)} e^{2\a 
(r)}d\Sigma_{n,\sigma}^2 + e^{2C(t)} e^{2\a (r)} (dx_0^2 + ... +dx_{k-1}^2)  \\ 
&+& e^{2E(t)} e^{2\theta (r)} (dy_1^2 + ... +dy_c^2) 
+ e^{2D(t)} e^{2\theta (r)} (dr^2+ r^2d\Omega _{m-c} ^2) \, . \nonumber
\eea
The p-brane is located at $(t, \Sigma_{n,\sigma}, x_0,...., x_{k-1})$. The  directions $(y_1, ..., y_c)$  are 
smeared for the p-brane which has $(m-c+1)$ dimensional localized transverse space. 
Here $\Sigma_{n,\sigma}$ is the intersection manifold with the S-brane which should be flat when the S-brane has charge. The radial metric functions are 
those of a p-brane, namely $e^{2\a}=H^{-\frac{d-p-3}{d-2}}$ and $e^{2\th}= H^{\frac{p+1}{d-2}}$ where the harmonic function $H$ is given 
before (\ref{harmonic}). With these choices field equations for radial and time-dependent parts decouple and one ends up with the usual S- and p-brane 
equations \cite{sp}. The number of intersection $n$ and smearing $c$ are fixed by solving the field equation for the non-diagonal Ricci tensor component:
\be
R_{tr}= (d-2-c) \a ' \dot{D}  
+ c  [\a'\dot{E}+ \th'(\dot{D}-\dot{E})] 
\,\, , \\
\label{mix}
\ee
where prime and dot correspond to $r$ and $t$ differentiations respectively.
Now  the S- and p-brane form fields in general do not contribute to the $R_{tr}$ component of the 
Einstein field equations (\ref{fieldeqns}), only the dilaton contributes \cite{sp}. Let us assume that there is no smearing, i.e. $c=0$, which 
needs to be the case if we want to keep the AdS near horizon geometry of a non-dilatonic p-brane, for which $a=0$, intact. 
From (\ref{mix}) it is clear that we should have $\dot{D}=0$. However, a time independent metric part in the general 
S-brane solution is not allowed when it is charged \cite{s2} but it is possible when it is chargeless \cite{deformations}.
Therefore, when S-brane has a charge, there has to be some smearing for a non-dilatonic p-brane. For example, in the M5-SM2 intersection 
we need $c=2$ or $c=1$ depending on whether radial part of M5-brane is included in the worldvolume of SM2 or not, respectively \cite{sp}.

General properties of these solutions related to time dependence are similar to usual S-branes \cite{s3, deger}.
First of all, note that one of the constants $\{M, \b, \g, b_i\}$ can be set to 1 by 
rescaling the time coordinate. There are generic singularities as $t \rightarrow \pm \infty$ as can be seen from the collapse of some of the metric 
functions. When $t$ is finite, time-dependent metric functions are well-behaved except the $G_{n,-1}$ function (\ref{G}) which becomes zero at $t=t_0$.
However, this is not a singularity; by defining a new coordinate $u = (t-t_0)^{-1/2}$ we get $-du^2 + u^2 \Sigma_{n,-1}$ which is just the 
flat spacetime in Rindler coordinates. Hence, for $\sigma =-1$ we have $t \in$ $(t_0, \infty)$ or  $(-\infty, t_0)$ whereas for $\sigma=\{0, 1\}$ we have
$t \in (-\infty, \infty)$. Finally, in the general solution (\ref{solution}) it is possible to smear some directions from the (m+1)-dimensional transverse space. 
For example, smearing one direction one has to do the replacement:
$H^{\frac{p+1}{d-2}} (dr^2+ r^2d\O_m^2) \rightarrow H^{\frac{p+1}{d-2}} (dz^2+ d\tilde{r}^2+ \tilde{r}^2d\O_{m-1}^2)$. The harmonic function $H$
is independent of the $z$-coordinate. Now, we would like to concentrate on the solutions for non-dilatonic p-branes, namely M2 and M5-branes in d=11 and, D3-brane 
in d=10, since they have AdS near horizon limits.

\subsection{Non-dilatonic p-branes}

For non-dilatonic p-branes we have $a=0$ and notice that in that case time dependence in the metric (\ref{solution}) appears only in the worldvolume directions. 
It is a well-known fact that one can replace the Minkowski geometry the worldvolume of a p-brane with any Ricci-flat spacetime \cite{ricci1, ricci2}. Hence, 
when also the constant $\gamma=0$ in (\ref{solution}), which is necessarily the case in d=11, our solutions just corresponds to this fact. But, otherwise the worlvolume 
replacement is not Ricci flat. 

One remarkable property of the solution that we found is that (\ref{solution}) it gives M5 and D3-branes in a time-dependent background.
However, M2-brane is not allowed. To see why, note that when the constant $\g=0$, the equation
(\ref{restriction}) implies that at least one of the 
$b_i$'s have to be nonzero, in other words $k \geq 1$. Combining this with the condition $n \geq 2$ this implies $p  \geq 3$ and hence
M2-brane solution in d=11 is not possible. From this argument it also follows that for the D3-brane 
$2 \leq n \leq 3$ and for the M5-brane $2 \leq n \leq 4$. 

Now let us focus 
on the near horizon $r \rightarrow 0$ region of the D3-brane. For $n=3$ this solution was obtained before for $\sigma=0$  in \cite{kasner1} and for $\sigma \neq 0$  
in \cite{frw1}. On the other hand, for $n=2$ only $\sigma=0$ case was considered before \cite{kasner1}.
It is obvious that when $\sigma=0$ both of these solutions approach to  $AdS_5 \times S^5$ geometry as $t \rightarrow t_0$.  
In this limit, also for $\sigma=-1$  the D3-worldvolume is conformal to parts of the Minkowski spacetime and hence is appropriate for studying time-dependent AdS/CFT 
correspondence too \cite{frw1}. For $n=3$ this follows immediately, since in this limit the D3 worldvolume becomes Rindler spacetime from the discussion above.
For $n=2$, the same argument works as well or one can see this more concretely by making change of variables\footnote{To put our metric into the form
used in \cite{frw1} one should define $e^{(n-1)M(t-t_0)}= |\tanh\frac{(n-1)\tau}{2}|$.} as in \cite{frw1}. There is an important difference between these 
two solutions in terms of the value of the constant $\gamma$, which plays a crucial role 
in the string coupling  $e^{\phi} =  e^{\gamma t}$. Notice that for $\sigma=-1$, 
$e^{\phi}$ is bounded unlike the $\sigma=0$ case and
as  we approach to the AdS boundary via $t \rightarrow t_0$, it becomes constant. 
Note that for $n=3$ we have $12M^2 = \gamma^2$ and for $n=2$ we have $4M^2 = 4\b^2 + \gamma^2$ from (\ref{restriction}).
Setting $M=1$ by rescaling the time coordinate, we see that 
$\gamma=2\sqrt{3}$ for $n=3$, whereas $\gamma$ can take any value from the interval [0,2] when $n=2$. As we approach to the singularity at 
$\t= - \infty$ in both $n=\{2,3\}$ cases the string coupling asymptotes to zero.
In \cite{frw2} it was argued that the singularity becomes better when the exponent in the string coupling is less than 1, which is possible only when $n=2$.

Similar arguments also work for the M5-brane where one gets
$AdS_7 \times S^4$ geometry for $\sigma=-1$ and $\sigma=0$ as $t \rightarrow t_0$. For M5-brane only the $\sigma=0$ case was known before \cite{kasner3}.


\section{Lunin-Maldacena Deformation of the Time Dependent M5-brane}

An important feature of S-branes is that they give rise to accelerating cosmologies upon compactification  \cite{townsend}.
When their charge is zero, as in the above solutions, this happens only when $\sigma=-1$, whereas for a charged S-brane there is no such 
restriction \cite{ohta}. Hence, it is desirable to find a solution with a charged S-brane which has an AdS limit to study accelaration from the
dual field theory point of view. To obtain such a solution, we recall a result of 
\cite{deform} where it was shown that applying a Lunin-Maldacena deformation \cite{lm} to a chargeless S-brane solution in d=11, one gets 
a charged SM2-brane. We now employ this idea to our time-dependent M5-brane solution.

In \cite{deformations} a simple formula was derived to obtain Lunin-Maldacena deformations \cite{lm}
of solutions of d=11 supergravity which possess three $U(1)$ isometries. The formula given in \cite{deformations} is valid only when these
$U(1)$ directions do not mix with any other coordinate in the metric 
and when the 4-form field strength has at most one overlapping with them. Since the metric of our M5-brane solution is 
diagonal, the first condition is automatically satisfied and hence we just need to be careful about the second one. Now, the 4-form field strength
of the M5-brane has directions in the transverse part and therefore we have only two possibilities for choosing the $U(1)$ directions: 
1 from transverse and 2 from the worldvolume or all of them from the worldvolume. Only the second choice gives rise to a new solution as we explain below.
For the first one, we need a smearing in the transverse space
which is always possible in (\ref{solution}) as explained before. For the M5-brane suppose we do this 
smearing and let there be at least two $y_i$ directions in (\ref{solution}). Then, using these two $y_i$'s and the smeared direction one gets 
a charged SM2-brane as was already observed in \cite{deform}. This is nothing but a generalization of the M5-SM2 solution found in \cite{sp} since we now allow
hyperbolic or spherical $\Sigma$'s on the worldvolume of the p-brane. Because of the smearing, there is no AdS limit.
For the second set of $U(1)$'s, let us now take the M5-brane without any smearing and choose $n=2$ in (\ref{solution}). We can now
use the remaining 3 worldvolume directions $\{y_0,y_1,y_2\}$ to perform Lunin -Maldacena deformation \cite{lm} which upon applying the method of \cite{deformations}
leads to:
\bea
ds^2 &=& H^{-\frac{1}{3}} \left[ K^{-\frac{1}{3}}e^{2 \b t} G_{2,\sigma}^{-1}(-G_{2,\sigma}^{-1} dt^2 + d \Sigma_{2,\sigma}^2) + 
 K^{\frac{2}{3}} \left(\sum_{i=0}^{2} e^{2 b_i t} dy_i^2\right)\right] \nonumber \\
\label{solutionM2M5}
&+& H^{\frac{2}{3}} (dr^2+ r^2d\O_4^2)  \,\,\,\, ,\\
\nonumber
\tilde{F}_{[4]} &=&  Q \, \textrm{Vol} (\O_4) - \a \, i_0i_1 i_2 * [Q\textrm{Vol} (\O_4)] + \a \, d (KH^{-1}e^{-2\beta t} dy_0 \wedge dy_1 \wedge dy_2) \, , \\
\nonumber
K &=& [1+\a^2H^{-1}e^{-2\beta t}]^{-1} \,\,\, , \,\,\, \beta=-(b_0 +b_1 + b_2) \, , 
\eea
where the Hodge dual * is with respect to the undeformed metric and $i_m$ stands for the contraction with respect
to the isometry direction $y_m$. Here $\a$ is the deformation parameter and when $\a=0$ we have the original time-dependent M5-brane solution.
From $\tilde{F}_{[4]}$ we see that there is an M2-brane located at $\{t,\Sigma_{2,\sigma}\}$ and a {\it static} SM2 located 
at $\{y_0,y_1,y_2\}$ whose 4-form field has a leg in $r$. When $\beta \neq 0$ we also have a
regular SM2-brane again at $\{y_0,y_1,y_2\}$. Note that all additional branes are entirely inside the M5-brane.
When all time dependence is removed this solution is just
the dyonic solution M2 $\subset$ M5 obtained in \cite{dyonic}. Note that in the near horizon limit $r \rightarrow 0$ for
any finite $t$ the function $K  \rightarrow 1$. Hence, we again have the $AdS_7 \times S^4$ geometry for $\sigma=-1$ and $\sigma=0$
as $t \rightarrow t_0$ and there is a cosmological singularity at infinity, like the single time-dependent M5-brane. Having an SM2-brane in the configuration is especially 
attracttive since compactifying on to its worldvolume results in an accelerating 4-dimensional spacetime \cite{townsend, ohta}.

\section{Discussion}

In this letter we have constructed a large class of new exact solutions in type $II$ and d=11 supergravities. Only some special cases of those corresponding to 
D3- and M5-branes were known before. However, their relation to S-branes was not realized. This opens up a whole new line of investigation 
since hyperbolic compactifications of chargeless S-branes produce a short period of accelerating cosmologies \cite{townsend}. Our time-dependent
M5-brane solution with $n=2$ and $\sigma=-1$ and the solution (\ref{solutionM2M5}) that we found in section 3 
are especially suitable to study this phenomena using the AdS/CFT duality.
There are of course many other aspects of this correspondence that one would like to understand better in time-dependent backgrounds 
and we expect our new solutions to be useful in studying them. For our solutions, understanding the decoupling of their 
worldvolume theories from the bulk \cite{ali} and studying the fate of cosmological 
singularity using the gauge theory picture are especially important.

As a side result, our work also shows that the S- and p-brane intersections found in \cite{sp} where the S-brane was charged,
can be generalized so that the p-brane worldvolume directions not intersecting with the S-brane
can be taken curved if their total dimension is greater than 1.

Intersections of the dyonic membrane solution \cite{dyonic} was studied in \cite{costa}. Repeating this for our time-dependent
version (\ref{solutionM2M5}) is desirable. Moreover, it is possible to add waves and Kaluza-Klein monopoles to p-branes \cite{kk}
and S-branes \cite{wave}. It would be interesting to investigate whether this is allowed for solutions found in this paper.

Studying black holes in a time-dependent AdS background is another very intriguing challenge.
Intersections of p-branes can be used to describe black holes in four and five dimensions \cite{bh1, bh2} which have AdS near horizons.
In \cite{wave} it was shown that adding a charged S-brane to these configurations is not possible. Nevertheless, our current work suggests that it is worth revisiting 
this question for a chargeless S-brane. We hope to investigate these problems in near future.

\

\noindent {\Large \bf Acknowledgements}

\noindent It is a great pleasure to thank Hermann Nicolai and Albert Einstein Institute for hospitality where this paper was written.
I also would like to thank Axel Kleinschmidt, Jean-Luc Lehners and Jan Rosseel for discussions. This work is partially supported by Tubitak grant 113F034.

\end{document}